# Origin of performance degradation in high-delithiation $Li_xCoO_2$: insights from direct atomic simulations using global neural network potentials

Pan Zhang,[1,2] Cheng Shang,[2,3] Zhipan Liu,[2,3] Ji-Hui Yang,[1,2*] and Xin-Gao Gong[1,2*]

[1]*Key Laboratory for Computational Physical Sciences (MOE), State Key Laboratory of Surface Physics, Department of Physics, Fudan University, Shanghai 200433, China*
[2]*Shanghai Qizhi Institution, Shanghai 200232, China*
[3]*Key Laboratory for Computational Physical Sciences (MOE), Shanghai Key Laboratory of MolecularCatalysis and Innovative Materials, Department of Chemistry, Fudan University, Shanghai 200433, China*

Email: jhyang04@fudan.edu.cn; xggong@fudan.edu.cn

## Abstract

$Li_xCoO_2$ based batteries have serious capacity degradation and safety issues when cycling at high-delithiation states but full and consistent mechanisms are still poorly understood. Herein, we provide direct theoretical understandings by performing long-time and large-size atomic simulations using the global neural network potential (GNNP) developed by ourselves. We propose a self-consistent picture as follows: (i) $CoO_2$ layers are easier to glide with longer distances at more highly delithiated states, resulting in structural transitions and structural inhomogeneity; (ii) at regions between different phases with different Li distributions due to gliding, local strains are induced and accumulate during cycling processes; (iii) accumulated strains cause the rupture of Li diffusion channels and result in the formation of oxygen dimers during cycling especially when Li has inhomogeneous distributions, leading to capacity degradations and safety issues. We find that large tensile strains combined with inhomogeneous distributions of Li ions play critical roles in the formation processes of blocked Li diffusion channels and the oxygen dimers at high-delithiation states, which could be the fundamental origins of capacity degradations and safety issues. While our molecular dynamics (MD) under strain effects provides direct evidence for the above findings, we note that the uniform distribution of Li ions can effectively improve the cyclicality and safety issues but is very challenging to realize especially at high-delithiation states. Correspondingly, a more practically feasible strategy is suppressing accumulations of strains by controlling charge and discharge conditions as well as suppressing the gliding, i.e., by inserting some strongly-bonded ions between the $CoO_2$ layers and/or by coating the $Li_xCoO_2$ grains, which will be helpful for improving the performance of lithium-ion batteries (LIBs). The current work demonstrates the feasibility and necessity of atomic simulations with a global perspective to provide more and critical insights into LIBs, thus opening new doors in this field.

# 1. Introduction

The past decades have witnessed rapid developments in portable electronics including mobile phones, laptops, wireless earphones, and electric vehicles, thanks to the technical revolutions of LIBs. As LIBs are playing an increasingly important role in our current society, high-performance LIBs in terms of high capacities, sufficient cycle times, and good safety are eagerly desired. To improve the capacities, high voltage operations are proposed to unlock the capacity potentials of current cathode materials including $LiCoO_2$[1-4], Ni-rich layered oxides[1, 5-8], $LiNi_xMn_yCo_zO_2$ (x+y+z=1)[1, 9-11], etc. For example, $LiCoO_2$ theoretically can reach a relatively high capacity (274 $mAhg^{-1}$) if all Li ions could be extracted under the high-voltage operation. However, in practical applications, only half Li ions can be extracted, resulting in the actual capacity of the $LiCoO_2$ being only 140 $mAhg^{-1}$.[4, 12] To utilize more Li ions, the operating voltage has to be above 4.2 V. However, in this case, the cycling performance plummets dramatically with serious capacity degradation and safety issues as reported in many experimental works.[4] Revealing the underlying mechanisms has become one of the most important issues in current research of LIBs, which is also essential for improving the actual performance and designing better electrode materials.

Among all possible reasons leading to capacity degradation, structural transitions at high-delithiation states are expected to play important roles. With the extraction of Li ions approaching the limit, experimental and theoretical works have reported that $Li_xCoO_2$ bulk transits from the O3 phase to the H1-3 phase for x<0.3 and then to the O1 phase at x=0.0.[13, 14] At the same time, many different experimental groups have reported capacity degradations due to various kinds of structural damages, like cracks[15-17], curved layers[18-20], dislocation[20-23], and even broken of $CoO_2$ layers.[20] While the transitions are attributed to the in-plane gliding of $CoO_2$ layers, the structural damages are owning to the local strains resulted from the gliding of $CoO_2$ layers. However, little is known about the gliding details, i.e., in terms of barriers, transition states, and gliding distances, not to mention how strains induce structural damages, especially at atomic levels. On the other hand, the release of oxygen at high-voltage cycling is expected to be responsible for safety issues. Previous works have tried to understand this issue from the lower $Co^{4+}/Co^{3+}$ redox level[24, 25], under-coordinated oxygen bondings[26-28], and formation energies of oxygen vacancies.[20] However, the atomic process of oxygen formation is rarely explored and only some first-principles MD simulations reported the release of $O_2$ at the O-terminated surfaces.[29, 30] To understand the capacity degradation and oxygen release at atomic levels and pave the theoretical gap between various observations and speculations, it is necessary to provide direct atomic simulations for the above dynamical processes in a large-scale space and long-time scale, which is, however, unfeasible currently, because of lack of powerful methods to treat such complex systems with both high efficiency and accuracy.

In this work, combining the neural network potential with the stochastic surface walking method (SSW-NN), we first develop a GNNP for $Li_xCoO_2$ systems which can accurately characterize the structural information and total energies for x ranging from

0.00 to 1.00, in good agreement with our first-principles calculations and the experimentally reported results. Then using the SSW-based response sampling (SSW-RS) approach, we explore the structural transitions of $Li_xCoO_2$. While all the previously reported structural transitions at various x values are reproduced, we also quantitatively identify the transition barriers and find that, $CoO_2$ layers are easier to glide with longer distances at higher-delithiation states. The consequence is that local strains can be easily induced due to gliding, especially at regions between different phases with inhomogeneous distributions of Li ions during the cycling. Considering the strain effects as well as the distributions of Li ions, we directly perform long-time MD (up to 8100 ps) for $Li_xCoO_2$ systems using our developed potentials. We find that large tensile strains exceeding 8% combined with inhomogeneous distributions of Li ions play critical roles in the formation processes of blocked Li diffusion channels and the oxygen dimers at high-delithiation states, which could be the fundamental origins of capacity degradations and safety issues. Consequently, to improve the performance of $Li_xCoO_2$ based batteries at high-delithiation states, it is necessary to improve the homogeneity of Li distributions during cycling and suppress the accumulations of local strains by controlling the charging and discharging conditions such as current, time, voltage, etc, and restraining the gliding of $CoO_2$ layers, i.e., by inserting some strongly-bonded ions between the $CoO_2$ layers and/or by coating the $Li_xCoO_2$ grains. Our simulations provide, for the first time, direct atomic insights into the full mechanisms of capacity degradations and safety issues.

## 2. Theoretical methods

The GNNP is trained using the SSW-NN method[31, 32] as implemented in the large-scale atomistic simulation with neural network potential (LASP) code.[33] To sample the global potential energy surface (PES) of $Li_xCoO_2$ at each specified composition, the SSW global optimization approach[34, 35] is adopted. To differentiate various structures located on the PES, the atomic structures are converted to power-type structural descriptors (PTSDs).[32, 36] Detailed density functional theory (DFT)[37, 38] and neural network (NN) settings can be found in the Supplementary Information S1. After a detailed functional test (see the Supplementary Information S2), we select the recently developed strongly constrained and appropriately normed (SCAN) density functional[39] to train the GNNP of $Li_xCoO_2$. The energies, forces, and stresses of the structures are then calculated using the DFT for training the neural network using the VASP program.[40, 41] The root-mean-square (RMS) errors for the energy, force, and stress on the training sets as a function of the training epoch can be seen in the Supplementary Information S3.1. Compared to the DFT results, the RMS errors for the energy, force, and stress of our trained GNNP are 8.015 meV per atom, 0.156 eV/Å, and 2.692 GPa respectively, indicating that the GNNP-PES is an excellent approximation to the DFT-PES of $Li_xCoO_2$. In addition, the energy-strain curves (see the Supplementary Information S3.2) at x=0.00, 0.50, and 1.00 using our GNNP agree very well with the DFT results over a large strain range (±30%), which demonstrate the high accuracy of our GNNP in describing the structural properties and total energies under strains. After obtaining the GNNP, the SSW-RS[42] method is used to search the critical structural transition pathways. Long-time and large-scale MD simulations are performed using LASP to study the atomic evolution processes under strains.

## 3. Result and discussion

### 3.1 Structural transition

The stoichiometric LiCoO$_2$ has an $\alpha$-NaFeO$_2$-type structure with a space group of $R\bar{3}m$ as shown in Fig. 1a, in which the oxygens are arranged in a cubic close-packed network while the Li$^+$ and Co$^{3+}$ ions are ordered in alternating (111) planes.[43] Such a stacking order of CoO$_2$ layers is denoted as the O3 phase. To mimic the delithiated states of Li$_x$CoO$_2$, nine compositions with x=1.00, 0.92, 0.83, 0.67, 0.50, 0.33, 0.17, 0.08, and 0.00 are considered. To identify the ground structures of each composition within the quasi-rock-salt framework, we start from the 4×4×1 supercell of LiCoO$_2$ in the O3 phase. By randomly removing Li ions, 30 structures are constructed with both the uniform and non-uniform distribution of lithium ions for each composition (except for x=1.00 and 0.00), which are then optimized by the GNNP. Then for each structure, we perform SSW-RS for 500 steps to search for final states with relatively low energies, which are also double-cross checked using the DFT to locate the ground states. The identified lowest energy structures for each composition are listed in Fig. S8 (see Supplementary Information S4.1) and Figs. 1b-1h. Based on these structures, we have calculated their formation energy and intercalation voltage profile, as provided in Fig. S11 (see Supplementary Information S4.2), which agrees well with previous studies. As we can see, at x>0.17 (see Fig. S8 and Figs. 1b-1d), Li$_x$CoO$_2$ always keeps the same O3 stacking order of CoO$_2$ layers as in the stoichiometric LiCoO$_2$ (see Fig. 1a). At x=0.17, the O3 and H1-3 phases are almost energetically degenerate (see Figs. 1e and 1f) with the O3 phase only 0.58 meV/atom lower. At high-delithiation states, the structures tend to have an H1-3 stacking type which consists of alternating blocks of the O1 and O3 stacking (see Fig. 1g). At a fully delithiated state with x=0.00, the lowest energy state totally becomes the O1 phase, as seen in Fig. 1h. These results agree with the experimental observations and theoretical prediction.[13, 14]

It's worth noting that, in the lowest energy structure at x=0.50 (see Fig. 1c and Fig. S12), the Li ions are ordered in a zigzag arrangement in which Li ions deviate from the center of the oxygen octahedron due to the Coulomb interactions between adjacent Li ions. The same structure is also reported using the Monte Carlo method[14] and the hybrid evolutionary algorithm.[44] Experimentally, the line mode arrangement of Li ions in Li$_{0.5}$CoO$_2$ is reported.[45] Our DFT calculations show that the zigzag arrangement is 2.48 meV/atom lower than the line mode arrangement. The discrepancy between theoretical and experimental results may be due to the dynamical behavior of Li ions under an external field at finite temperature.

At x=0.17 (see Fig. 1f), a Frenkel-type defect consisting of a Co interstitial at the octahedron site in the Li-ion layer and a subsequent Co vacancy at the octahedron site in the CoO$_2$ layer, is formed, which is consistent with the experimental observations of Co diffusing into Li sites. At x=0.08 (see Fig. 1g), the structure completely transits to the H1-3 phase with the Frenkel defect kept. At a fully delithiated state with x=0.00, the lowest energy structure totally becomes an O1 phase, as seen in Fig. 1h. The Frenkel defect is still present as our DFT results show that this structure is 1.27 meV/atom lower than that without the Frenkel defect.

In addition to the stacking orders of the $CoO_2$ layers, we also examine the dependence of lattice constants on the lithium concentrations. As shown in Fig. 2a, with the desertion of Li ions, the projection of lattice constant $c$ along the $z$-direction $c_z$ first increases, reaches the maximum at x=0.50, and then rapidly decreases, undergoing structural transitions from the O3 stacking to the H1-3 and then to the O1 stacking. Compared to $c_z$, the in-plane lattice constant $a$ changes oppositely with a relatively smaller range as seen in Fig. 2b, in good agreement with experimental observations. Note that, the large reduction of $c_z$ at high-delithiation states is expected to be responsible for the reduced device performance as it will be more difficult for Li ions to diffuse in this case. We note that the changes of $c_z$ are strongly related to the stacking of $CoO_2$ layers, especially at high-delithiation states (see Fig. S13 in the Supplementary Information S4.4). For example, at x=0.17 with the H1-3 phase (see Figs. S13c-S13d), there are two layers with the O1 stacking type. At x=0.08 (see Figs. S13e-S13h), three O1-stacking layers show up. At x=0.00 (see Figs. S13i-S13j), all the layers are stacked in the O1 type and $c_z$ reaches the minimum. Consequently, suppressing the O1 stacking at high-delithiation states might be crucial for LIB performance.

To reveal how O1 stacking arises with the delithiation, we turn to study the structural transitions using the SSW-RS methods. We utilize the variable-cell double-ended surface walking (DESW) method[46] to establish the pseudo-pathway which connects the initial state to the final state for all initial/final state pairs.[47, 48] The approximate barrier is obtained according to DESW pseudo-pathway, where the maximum energy point along the pathway is generally a good estimation for the true transition state.[46] For each composition except for x=1.00 and 0.00, we construct 30 initial O3 structures which have different Li arrangements. For each structure, we consider about 500 transition paths and thus about 15000 transition paths are considered for each composition (except that for x=1.00 or 0.00 only one initial structure is constructed and about 5000 transition paths are considered). For each composition, we finally collect 300 paths which lead to the final states with relatively low energies. We note that the structural transitions are mainly caused by the relative gliding of neighboring $CoO_2$ layers, which can be roughly divided into three regions according to the emergence of the new types of stacking sequences of $CoO_2$ layers, which are 0.33≤x≤1.0, 0.00<x<0.33, and x=0.00, respectively. In the first region, both the initial and final structures keep the O3 stacking with the gliding of $CoO_2$ layers. Such gliding which happens before the deep delithiation is rarely reported experimentally. In the second region, due to the gliding of $CoO_2$ layers, the H1-3 stacking could show up in the final structures. In the third region, the final structures could have the H1-3 or O1 stacking. The energy barriers versus gliding distances of $CoO_2$ layers are shown in Fig. 3, in which the gliding distance of zero only means the hopping of discrete Li or Co ions but not real structural transitions. Generally speaking, with the decrease of x, the structural transition barriers become lower while the gliding distances get longer. For example, at x=0.67, the closest gliding distance is 1.440 Å/Co ion with a barrier of 0.735 eV and the farthest gliding distance is 8.514 Å/Co ion with a rather high barrier of 3.293 eV. At x=0.08, the smallest and largest gliding distances are 1.400 Å/Co ion and 8.396 Å/Co ion, respectively, with transition energy barriers of 0.409 eV and 1.625 eV, respectively. At x=0.00, the closest gliding distance is 2.015 Å/Co ion with a barrier of 0.540 eV and the farthest gliding distance reaches 9.142 Å/Co ion with a barrier of only 1.368 eV.

Note that, the structural transitions are accompanied by the migrations of Li ions. Several typical transition paths as well as the transition states are provided in the Supplementary Information S4.5 and S4.6, where we use the DESW transition-state-search method[46] to exactly locate the "true" transition state of the above transition pathways. We find that, in the first region, such as x=0.67, the Li ions in the transition state are in a tilted LiO triangular prism due to the slip of the $CoO_2$ layer (see Fig. S15a), which is different from the case of one Li ion diffusion in the O3 phase without the gliding of $CoO_2$ layers.[49] At high-delithiation states, the gliding of $CoO_2$ layers is always accompanied by the migration of Co ions. In the Supplementary Information S4.6, we construct an ideal model without the gliding of $CoO_2$ layers to study the formation process of the migration of Co ions and find that the diffusion of Li ions in one layer makes it easy for Co ions to diffuse into the Li-ion layer, which is the key step in the experimentally widely observed formation of the spinel phase.[45, 50, 51]

Here we make some discussions about the consequences of the gliding of $CoO_2$ layers. Such changes of stacking sequences have been widely observed experimentally in common Li- and Na-ion battery materials during the (de)lithiation.[1, 13, 14, 52-58] If the crystalline of $Li_xCoO_2$ was very homogenous and the removal of Li ions was uniform, the easy layer gliding, especially at high-delithiation states, was unlikely to cause serious structural problems, because in the gliding process, each layer is topotactically preserved and only the relative alignments between neighboring layers change. Besides, we also consider the electronic structures of the transition states and no significant differences are found in the density of states, indicating that the gliding itself is not likely to be directly responsible for the release of $O_2$. However, in the practice, the reactions are often inhomogeneous, leading to various areas with different distributions of Li ions.[4] Consequently, the gliding of $CoO_2$ layers is also not uniform, which is very likely to result in deformation and stress accumulation at the local interfaces between different areas, especially along the *ab*-plane at high-delithiation states because of the much lower gliding barriers. In fact, structural damages such as the formation of microcracks, accumulation of dislocations, accumulation of extended defects, and the changes in the shape of active-material particles, have been widely observed experimentally.[1, 22, 50, 59, 60] However, very little is known about the dynamic process and details, i.e., what extent of strains can induce structural damages, which will be studied in the followings.

**3.2 Lattice distortion caused by strain**

To study the strain effects on LIBs, we perform long-time MD simulations under strains for different delithiated states (x=1.00, 0.50, 0.33, 0.00) using our GNNP. Both the cases of uniform and non-uniform distributions of Li ions are considered without bias. We note that, the edges of $Li_xCoO_2$ are dominated by the [104] and [012] planes[29, 30] and therefore $CoO_2$ layers are likely to glide along the $b$-direction shown in Figs. 4a. In the following discussion, for simplicity we just focus on the strains along the $b$-axis shown in Figs. 4a and 4b and we expect similar results should be obtained if strains are along the other directions. To identify what extent of strains can cause structural damages, we gradually increase strains step by step and at each step, sufficiently long MD simulations are performed to make sure the systems are equilibrium. Details are shown in the followings.

First, we consider the effects of pure compressive strains on the structures of $Li_xCoO_2$. We apply a compressive strain along the $b$-axis which is gradually increased from 0% with a step size of -1%. At each step, we perform 100 ps MD simulations using an NVT ensemble (T=300K). The MD time step is 1fs. For each strain step, a longer MD time of 1000 ps was also tested and no significant changes were found. For x=1.00, the distribution of Li ions is uniform and we find the system does not show any apparent changes even under -8% strains except that the Co-O bond lengths are changed to fit the stress caused by the strain (see Fig. S18c). For x=0.50 and x=0.33, similar situations as the case of x=1.00 are found when the distribution of Li ions is uniform (see Fig. S19c and Fig. S21c). However, when the distribution of Li ions is non-uniform, curved layers appear at the -8% strain in the region with a relatively sparse Li-ion concentration (see Fig. S20c and Fig. 4c) while the area with relatively dense Li ions still maintains the initial arrangement without much change. For the case of x=0.00 (see Fig. S22c), it is similar to the cases when the Li ions are uniformly distributed. Note that, for all the delithiated states considered here, even under the large compressive strain of -8%, the Li-ion channels are not broken or blocked. In fact, if we remove the compressive strain, the system will return back without any structural damages, indicating the compressive strain itself is not irreversibly harmful to the performance of LIBs.

Using x=0.33 with the non-uniform distribution of Li ions as a typical example, the detailed simulation results are given in Fig. S23 in the Supplementary Information S5.2. As can be seen that, when the compressive strain is gradually increased from 0% to -5% (see Figs. S23a-S23f), the system does not show any apparent changes except that the Co-O bond lengths are changed to fit the stress caused by the strain. When the strain is increased to -6% (see Fig. S23g), curved layers appear in the regions with a relatively sparse Li-ion concentration. In contrast, the area with relatively dense Li ions still maintains the initial arrangement without much change. When the strain is further increased from -6% to -8% (see Figs. S23h-S23i), the whole system is subjected to a curved arrangement (see Fig. 4c). These results can be understood as follows. When the system is subjected to a small degree of compressive strain, the stress can be released by changing the bond lengths. However, when the system is subjected to a more significant degree of compressive strain, the system will have to release this part of strain by changing the bond angles, thus forming the curved layers commonly observed in our simulations and experiments.[19, 20]

Next, we consider the effects of pure tensile strains on the structures of $Li_xCoO_2$. Similar to the case of compressive strain, we apply a tensile strain along the $b$-axis, gradually increase it from 0% with a step size of 1%, and perform 100 ps MD simulations using an NVT ensemble for each strain step. A longer MD time was also tested for each strain step and no significant changes were found. In general, when the distribution of Li ions is uniform including x=1.00 and x=0.00, the systems do not show any apparent changes even under 8% strains except that the Co-O bond lengths are changed to fit the stress caused by the strains (see Figs. S18d, S19d, S21d, and S22d for x=1.00, 0.50, 0.33, and 0.00, respectively). However, for the cases when the distribution of Li ions is non-uniform, with the tensile strain gradually increased to 8%, the original Li-ion migration channels are blocked and new channels are formed due to the fracture of the $CoO_2$ layers (see Fig. S20d for x=0.50 and Fig. 4d for x=0.33). Note that, if we remove the tensile strain, the structural changes will not vanish, indicating the large tensile strains will cause irreversible damages to the performance of LIBs.

Using x=0.33 with the non-uniform distribution of Li ions as a typical example, the detailed simulation results are given in Fig. S24 in the Supplementary Information S5.3. Same as the case of the compressive strain, when tensile strain is less than 6% (see Figs. S24a-S24g), the system adapts to the applied stress mainly by changing the Co-O bond lengths. When the stress is increased to 7% (see Fig. S24h), a local defect structure shows up in which the CoO octahedron is stretched into a CoO quadrilateral in the region with relatively sparse Li ions. Further increase of the tensile strains to 8% leads to the blockage of the original Li-ion migration channels as well as the formation of new channels and the formation of oxygen dimers marked by a red dotted box due to the fracture of the $CoO_2$ layers (see Figs. 4d and S24i). Here, we make a detailed discussion about the MD process under tensile strains up to 8% (see Fig. S25 in the Supplementary Information S5.4). The initial structure is from the simulation result of 7% strain with the CoO octahedron stretched into a CoO quadrilateral in the region with relatively sparse Li ions, as seen in Fig. S25a. Then CoO quadrilaterals gradually appear in every $CoO_2$ layer (see Figs. S25b-S25c). With further thermal motions, these CoO quadrilaterals are reconnected, forming a nucleus of the rutile phase[44, 61] and blocking channels of Li ions (see Fig. S25d). Subsequently, a new CoO quadrilateral is formed (see Fig. S25e), making the two interlayer oxygen ions around the fracture closer (see 1.424 Å) and forming an oxygen dimer (see Fig. S25f). Then a new and more stable $CoO_2$ octahedron forms on the left side of the oxygen dimer (see Fig. S25g). Further thermal motions of ions lead to the breakdown of some $CoO_2$ layers (see Fig. S25h). Note that, different from the case of compressive strains, the crack formed under tensile strains can't be irreversibly eliminated and the system will not return back even if the tensile strain is removed.

The fact that the regions with different Li-ion concentration have different behaviors under strains can be understood from the stress-strain curves[62] which clearly show that under the same stress, the fully delithiated $CoO_2$ (corresponding to the extreme Li sparse case) will produce more significant strain compared with the fully lithiated $LiCoO_2$ (corresponding to the extreme Li dense case). This can also explain why the region with relatively sparse Li ions is more prone to form curved layers and fracture.

From the above studies of the cases with just pure compressive or tensile strains, we notice that high-deliciated $Li_xCoO_2$ can sustain -8% strain but show irreversible structural damages under 8% strain when Li ions have ununiform distributions. Since the large strains could result from an accumulation process during the practical charging and discharging cycling with Li ions in and out, we further mimic such situations by applying alternative compressive and tensile strains on $Li_xCoO_2$ (x=1.00, 0.50, 0.33, 0.00) along the *b*-direction for simplicity, i.e., from 0% to -5% and then from -5% to 5% and then back to 0% with a step of 1%. For each strain step, we perform 100 ps MD simulations in an NVT ensemble (T=300K). The above procedures go on and on for 5 cycles. Our results show that small strains like 5% do not cause structural damages but large strains like 8% will. In the following, we mainly present the case of large strains, i.e., from 0% to -8% and then from -8% to 8% and then back to 0% with a step of 2%. The procedures go on and on for 5 cycles and the total simulation lasts 8100 ps. The applied alternative strains as a function of the time can be seen in Fig. 5a.

Our simulation results show that, in the cases with uniform distribution of Li ions, alternative strains do not cause apparent structural damages (see Figs. S26, S27, S29, and S30 for x=1.00, 0.50, 0.33, and 0.00, respectively). However, when the distribution of Li ions is non-uniform, the alternative strains lead to the blockage of the original Li-ion migration channels in the regions with a relatively sparse Li-ion concentration due to the fracture of the $CoO_2$ layers (see Fig. S28 for x=0.50 and Figs. 5b-5g for x=0.33).

Using x=0.33 as a typical example, we show the detailed evolution processes. Figs. 5b-5g shows the snapshots denoted by the red points shown in Fig. 5a to demonstrate the effect of alternative strain along *b*-direction on the structures of $Li_{0.33}CoO_2$ with the non-uniform distribution of Li ions. While the system keeps the initial structure in the first 100 ps under zero strain (see Fig. 5b and Fig. S31a), it starts to get curved under compressive strains (see Fig. S31b). Nevertheless, the system restores back when the compressive strains are removed (see Fig. S31c). However, under tensile strains, some neighboring $CoO_2$ layers get broken and reconnected, which block Li-ion migration channels (see Fig. S31d). What's more, these $CoO_2$ layers remain connected after the tensile strain is gradually eliminated (see Fig.5c and Fig. S31e). Now the first cycle is finished and the structure is similar to the case of tensile strains. In the following two cycles (see Figs. 5c-5e and S31e-S31o), the connecting $CoO_2$ layers gradually evolve with Co-O bonds broken and reformed. With the simulation time increasing, oxygen dimers show up at the connecting ends (see Figs. 5f-5g and S31p-S31x), which might be owing to the much larger strains accumulated in this area during the cycling.

We have repeated the above simulations several times and the results are summarized in the Table 1. As can be seen, the uniform distribution of lithium ions during the charging and discharging cycling is crucial for the stability and safety of LIBs. For most cases, no broken of $CoO_2$ layers or formation of oxygen dimers show up except for the case of x=0.33 with uniform distribution of Li ions in group 3. Detailed analysis shows that (see Fig. S32 in the Supplementary Information S5.7), at high-delithiated states like x=0.33, the migration of Li ions is very easy to cause the distribution of Li ions deviating from homogeneity, resulting in the subsequent broken of $CoO_2$ layers and formation of oxygen dimers during cycling. One more thing worth mentioning is that, for x=0.33 with non-uniform distribution of Li ions in group 3, we find that dislocations of $CoO_2$ layers will show up and trap Li ions (see Fig. S33), in agreement with experimental observations.[63] Due to the presence of these dislocations, the chemical-mechanical stability of the system is kind of improved, as no new cracks formed in the subsequent cycles (see Figs. S33e-S33f). In this sense, the dislocation defect, which is typically regarded as harmful, accommodates the tensile stress to some extent and thereby delays the mechanical degradation.[64] Besides, we find an interesting phenomenon that, although both oxygen dimers and dislocations have been identified in different simulations, they seem not to appear in stable forms simultaneously. This is understandable since the formation of oxygen dimers requires large local strains while dislocations release strains.

With all the above simulation results in hand, now we can, for the first time, understand the capacity degradations and safety issues directly from the atomic level. With the extraction of Li ions, it's easier for the $CoO_2$ layers to glide with relatively lower barriers but longer distances. Due to the inhomogeneous distribution of Li ions and different structural phases of $Li_xCoO_2$ especially at high-delithiation states, the gliding of $CoO_2$ layers causes local strains, which could accumulate and evolve during the cycling. When the accumulated strain is large enough, i.e., exceeding 8%, the strains will cause the broken and reformation of Co-O bonds together with the non-uniform distribution of Li ions, resulting in the rearrangements of ions and formation of reconnected $CoO_2$ layers. As a result, some Li ions are trapped and Li-ion migration channels are blocked, leading to irreversible capacity degradation. Besides, the formation of oxygen dimers could be the initial origin of oxygen release, especially when this happens close to the surface with cracks of $Li_xCoO_2$ particles. Therefore, we suggest that experimental scientists can provide an excellent control scheme to slow down the accumulation of strains in the electrode materials by studying various variables during the charging and discharging process, such as current, voltage, temperature, etc. Besides, suppressing the gliding, i.e., by inserting some strongly-bonded ions between the $CoO_2$ layers and/or by coating the $Li_xCoO_2$ grains which are practically feasible, will also be helpful for slowing down the accumulation of strains. More dynamical simulations for the more complicated surfaces and interfaces will be studied in future works.

## 4. Conclusion

In conclusion, using the GNNP developed by ourselves in this work, we have provided direct theoretical understandings by performing long-time and large-size atomic simulations. We have found that $CoO_2$ layers are easier to glide with longer distances at higher-delithiation states, resulting in structural transitions and structural inhomogeneity. Due to the gliding, local strains can be induced at regions between different phases with different Li distributions and could accumulate during cycling processes. Our MD simulations under strains have shown that accumulated strains exceeding 8% could cause the rupture of Li diffusion channels and result in formation of oxygen dimers during cycling especially when Li has inhomogeneous distributions, leading to capacity degradations and safety issues. We have found that the uniform distribution of Li ions plays crucial roles for improving the cyclicality and safety issues.

Nevertheless, it is very challenging to maintain the homogeneity of Li ions especially at high-delithiation states. Correspondingly, suppressing the accumulation of strains in the electrode materials by studying various variables during the charging and discharging process, such as current, voltage, temperature, etc., will be beneficial. Besides, suppressing the gliding, i.e., by inserting some strongly-bonded ions between the $CoO_2$ layers and/or by coating the $Li_xCoO_2$ grains which are practically feasible, will also be helpful for slowing down the accumulation of strains. Our work demonstrates the feasibility and necessity of atomic simulations to provide more and critical insights into LIBs for the performance optimization.

## Acknowledgment

This work was supported in part by the National Natural Science Foundation of China Grant No. 12188101, No.11991061, No.11974078, and No. 61904035 and partly by the funding from the CATL. The trainings of our GNNP as well as all the computations were performed using the resources at the High-Performance Computing Center of Fudan University.

# Figures and tables

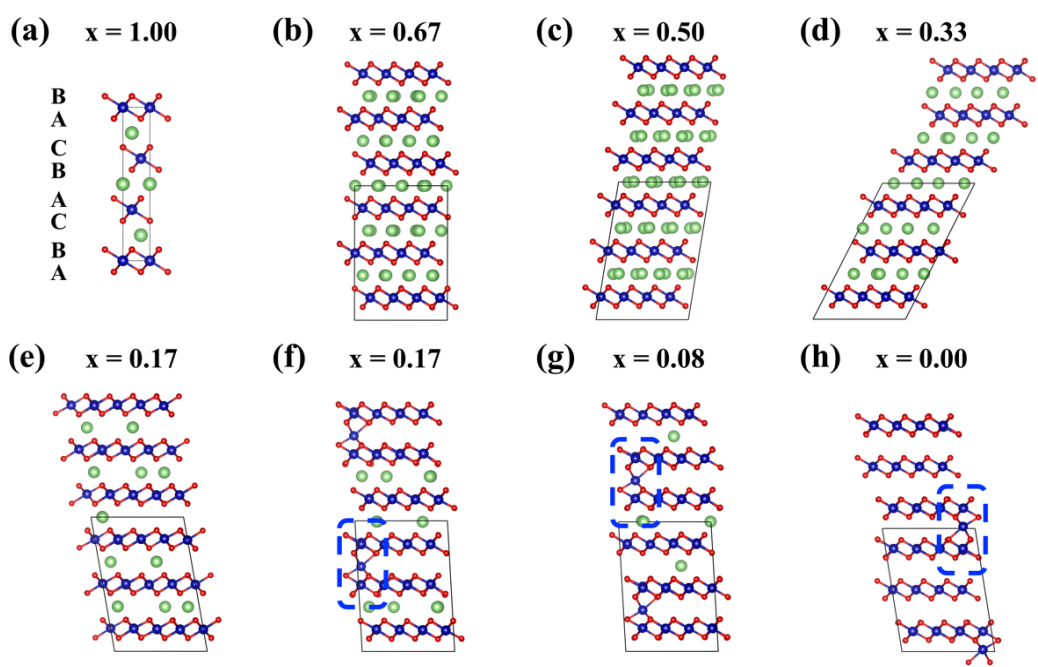

Figure 1. Lowest energy structures of $Li_xCoO_2$ identified using the SSW-RS method. (a) The stoichiometry $LiCoO_2$ structure with the O3 stacking type. (b) $Li_{0.67}CoO_2$, (c) $Li_{0.50}CoO_2$, (d) $Li_{0.33}CoO_2$, and (e) $Li_{0.17}CoO_2$ with O3 stacking type. (f) $Li_{0.17}CoO_2$ and (g) $Li_{0.08}CoO_2$ with H1-3 stacking type. (h) $CoO_2$ with O1 stacking type. Blue dotted boxes are used to mark local defects related to Co ions. Li, Co, and O are shown in green, blue, and red, respectively.

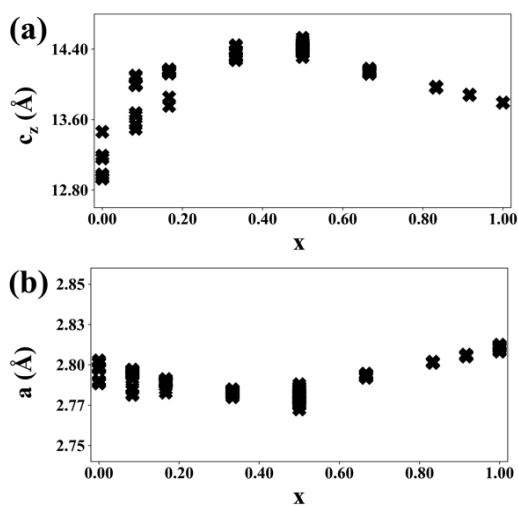

Figure 2. Lattice constants as functions of x in $Li_xCoO_2$. (a) The projection of lattice constant $c$ along the $z$-direction $c_z$. (b) The in-plane lattice constant $a$ of $Li_xCoO_2$. For each x, we collect 50 structural configurations with the lowest energies.

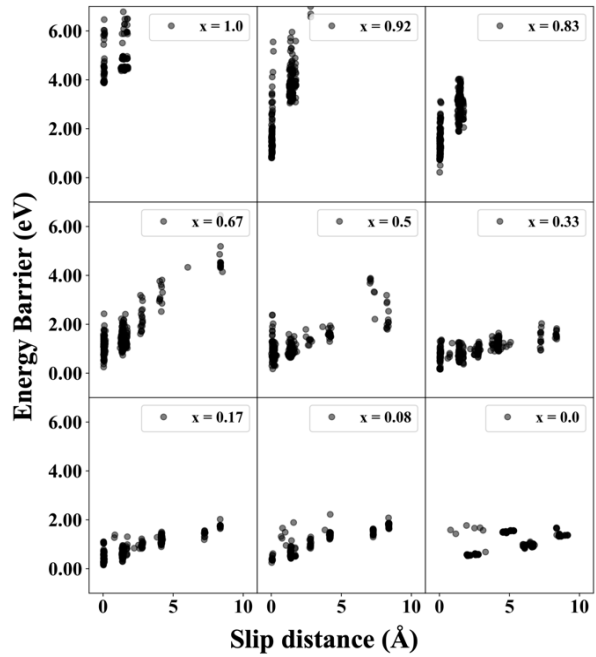

Figure 3. The relationship between the gliding distances of the Co-ions per Co ion along the *ab*-plane and the barriers to be crossed during the structural transitions. Note that, with the de-lithiation of Li ions, the system is more likely to undergo longer distances with a lower barrier.

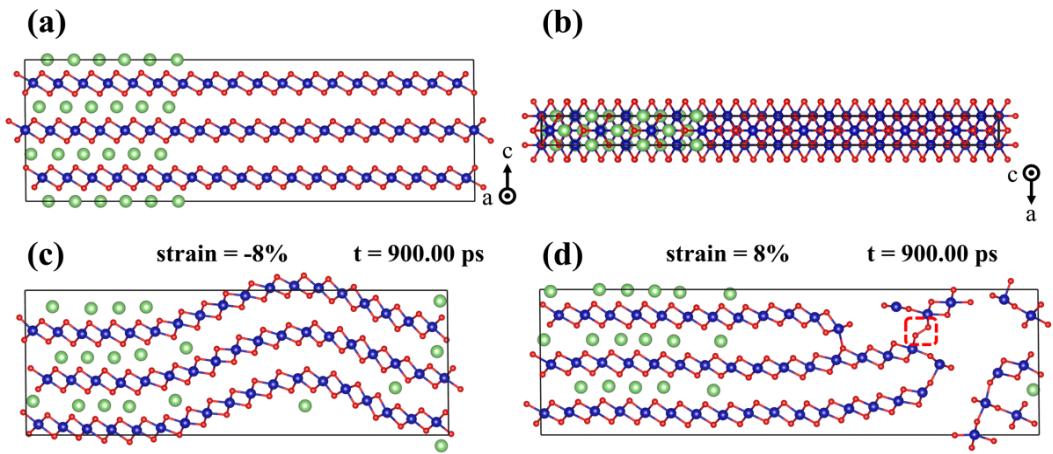

Figure 4. Structures of $Li_{0.33}CoO_2$ with the non-uniform distribution of Li ions under zero, compressive, and tensile strains. (a) Front and (b) top view of initial $Li_{0.33}CoO_2$ supercells under zero strains. (c) Structure of $Li_{0.33}CoO_2$ under a -8% compressive strain after 900 ps MD simulations at 300 K. (d) Structure of $Li_{0.33}CoO_2$ under an 8% tensile strain after 900 ps MD simulations at 300 K. Li, Co, and O are shown in green, blue, and red respectively. When the compressive strain is applied, curved layers appear in the regions with a relatively sparse Li-ion concentration. When the tensile strain is applied, the blockage of the original Li-ion migration channels, the formation of new channels, and the formation of oxygen dimers can be directly observed due to the fracture of the $CoO_2$ layers.

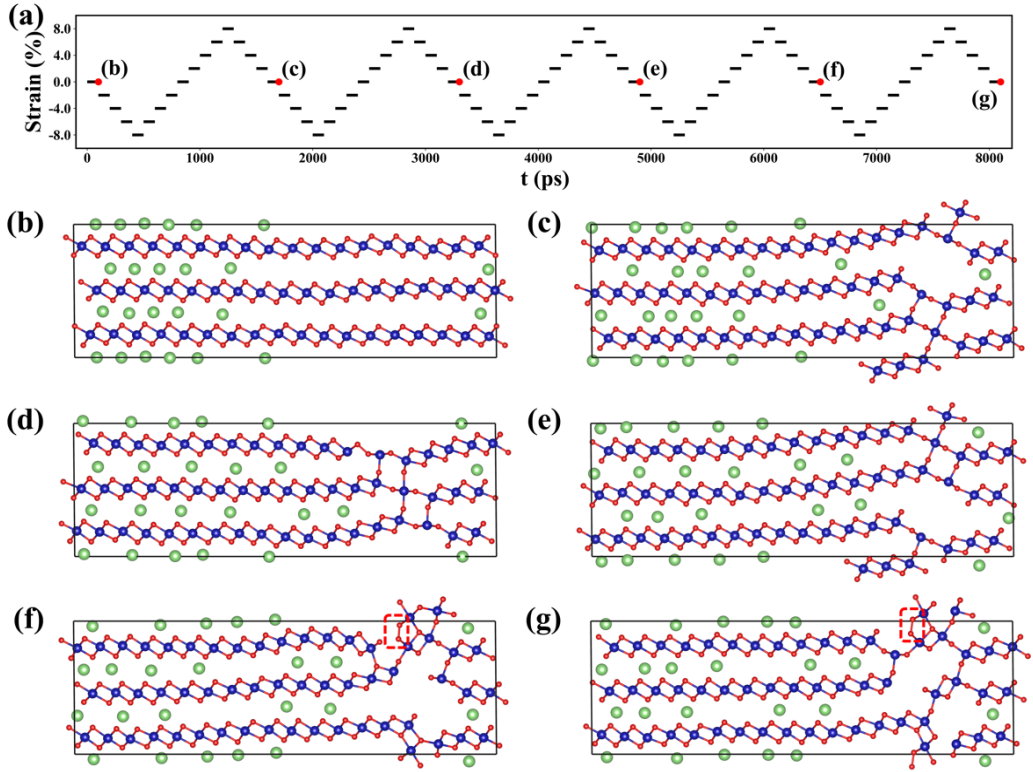

Figure 5. (a) The applied alternative strains as a function of the time. (b)-(g) The snapshots denoted by the red points shown in (a) to demonstrate the effect of alternative strain along *b*-direction on the structures of $Li_{0.33}CoO_2$ with the non-uniform distribution of Li ions. Li, Co, and O are shown in green, blue, and red respectively. Note that, the oxygen dimers arise at the end of our long-time MD simulations.

Table 1. The results of repeated tests for different delithiation states (x=1.00, 0.50, 0.33, 0.00) under alternative strains with uniform and non-uniform distribution of Li ions.

| x | Group number | Uniform distribution | | | | Non-uniform distribution | | | |
|---|---|---|---|---|---|---|---|---|---|
| | | Fracture caused by tensile strain | Curving caused by compression strain | The broken of $CoO_2$ layer casued by cycles | The formation of O-dimer caused by cycles | Fracture caused by tensile strain | Curving caused by compression strain | The broken of $CoO_2$ layer casued by cycles | The formation of O-dimer caused by cycles |
| 1.00 | 1 | No | No | No | No | \ | \ | \ | \ |
| | 2 | No | No | No | No | \ | \ | \ | \ |
| | 3 | No | No | No | No | \ | \ | \ | \ |
| 0.50 | 1 | No | No | No | No | Yes | Yes | Yes | No |
| | 2 | No | No | No | No | Yes | Yes | Yes | No |
| | 3 | No | No | No | No | Yes | Yes | Yes | Yes |
| 0.33 | 1 | No | No | No | No | Yes | Yes | Yes | Yes |
| | 2 | No | Yes | No | No | Yes | Yes | Yes | Yes |
| | 3 | No | Yes | Yes | Yes | Yes | Yes | Yes | No |
| 0.00 | 1 | No | No | No | No | \ | \ | \ | \ |
| | 2 | No | No | No | No | \ | \ | \ | \ |
| | 3 | No | No | No | No | \ | \ | \ | \ |